\documentclass{PoS}

\usepackage{amssymb}
\usepackage{times}
\usepackage{mathptmx}
\usepackage{fontenc}
\usepackage{psfig}
\usepackage{graphicx}
\usepackage{cite}

\title{Observationally-Motivated Analysis of Simulated Galaxies}

\ShortTitle{Observationally-Motivated Analysis of Simulated Galaxies}

\author{\speaker{Maider S. Miranda}
\thanks{Special thanks to our collaborators S. Pasetto, D. Kawata, 
C. Brook, G. Stinson, and C. Few.}\\
University of Central Lancashire\\
E-mail: \email{msancho@uclan.ac.uk}}

\author{Ben A. MacFarlane\\
University of Central Lancashire\\
E-mail: \email{bmacfarlane@uclan.ac.uk}}

\author{Brad K. Gibson\\
University of Central Lancashire\\
E-mail: \email{brad.k.gibson@gmail.com}}

\abstract{The spatial and temporal relationships between stellar age, 
kinematics, and chemistry are a fundamental tool for uncovering the 
physics driving galaxy formation and evolution. Observationally, these 
trends are derived using carefully selected samples isolated via the 
application of appropriate magnitude, colour, and gravity selection 
functions of individual stars; conversely, the analysis of 
chemodynamical simulations of galaxies has traditionally been restricted 
to the age, metallicity, and kinematics of `composite' stellar particles 
comprised of open cluster-mass simple stellar populations.  As we enter 
the Gaia era, it is crucial that this approach changes, with simulations 
confronting data in a manner which better mimics the methodology employed 
by observers. Here, we use the \textsc{SynCMD} synthetic stellar populations tool 
to analyse the metallicity distribution function of a Milky Way-like 
simulated galaxy, employing an apparent magnitude plus gravity selection 
function similar to that employed by the RAdial Velocity Experiment 
(RAVE); we compare such an observationally-motivated approach with that 
traditionally adopted - i.e., spatial cuts alone - in order to 
illustrate the point that how one analyses a simulation can be, in some 
cases, just as important as the underlying sub-grid physics employed.}

\FullConference{XIII Nuclei in the Cosmos\\
7-11 July, 2014\\
Debrecen, Hungary}

\begin{document}

\section{Introduction}

\begin{figure}[h]
\begin{center}
\hspace{0.25cm}
\psfig{figure=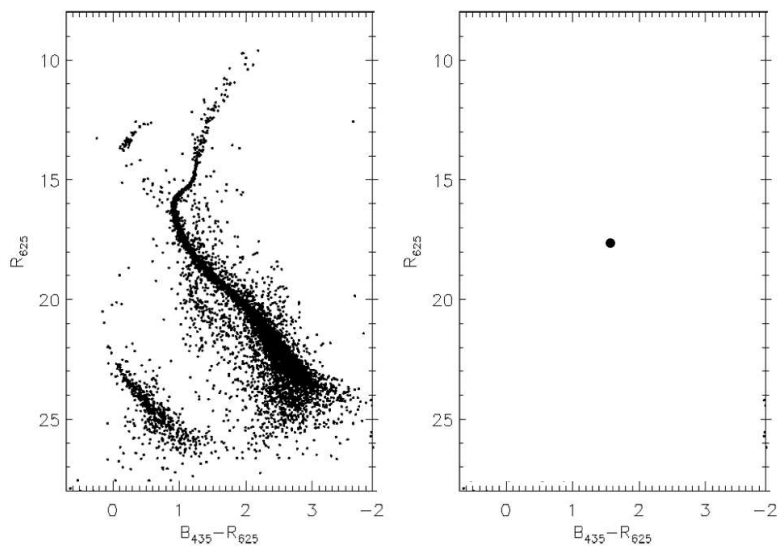,width=11.cm}
\caption{Left panel: A real colour-magnitude diagram comprised of 
$\sim$10$^4$ stars drawn from NGC~6397 \cite{Strickler09}. Right panel: 
How the same $\sim$10$^4$ stars are represented by `composite' stellar 
particles within a simulation.}
\label{cmdobs}
\end{center}
\end{figure}

An inherent assumption underpinning the analysis of galaxy simulations 
is that they provide an analogous framework to that of galaxies in 
nature, in the sense that the chemo-kinetics of the massive stellar-like 
particles in simulations can be compared directly with the individual 
stars in observational samples. Considering that the typical mass of the 
`composite' stellar particles in cosmological, hydrodynamical, 
simulations of Milky Way-like galaxies is of order 
$\sim$10$^{~5\pm1}$~M$_\odot$, rather than the 
$\sim$10$^{0\pm1}$~M$_\odot$ encountered in nature, it is not unreasonable 
to question the importance of this mismatch in `resolution'; 
graphically, this `mismatch' can perhaps best be appreciated by 
referring to Fig~\ref{cmdobs}.  Further, observers construct samples of 
stars based upon their respective and individual colour, apparent 
magnitude, chemistry, and gravity, while simulators are limited 
primarily to the integrated luminosities and chemistry of the coeval, 
open cluster-scale, simple stellar populations used to represent the 
`composite' stellar particles.

In what follows, we present a pilot study for our long-term programme 
aimed at developing powerful, and flexible, tools for better 
transforming simulations into the observer's plane. Here, we employ the 
synthetic colour-magnitude diagram package \textsc{SynCMD} 
\cite{Pasetto12} to replace composite stellar particles with individual 
stars in a self-consistent fashion, thereby allowing stellar samples to 
be drawn by an observer situated within a simulation using apparent 
magnitude and gravity criteria. The impact of this 
observationally-motivated selection function upon the inferred 
metallicity distribution of an analogous `solar neighbourhood', compared 
with the traditional approach (i.e., selecting composite stellar 
particles spatially), will be assessed, and our future directions 
outlined.

\section{Methodology}

The characteristics of the $\sim$L$\ast$ Milky Way-analogue 
analysed for this work (MaGICC-g1536) are detailed in \cite{Gibson13}.  
The simulation was realised with the smoothed particle hydrodynamics 
code \textsc{Gasoline} \cite{Wadsley04}, employing the MaGICC (Making 
Galaxies in a Cosmological Context) stellar feedback prescription 
\cite{Brook12}.
To transform the simulation into the observer's plane, we use the 
\textsc{SynCMD} (synthetic colour-magnitude diagram tool) developed by 
\cite{Pasetto12}. Using the age, metallicity, and initial mass function 
underpinning stars in simulations, 
\textsc{SynCMD} decomposes each particle into its individual constituent 
(or `synthetic') stars (of order $10^{~5}$ `synthetic' stars per 
`composite' stellar particle, corresponding to $\sim$10$^{~11}$ 
individual stars for MaGICC-g1536), by mapping the composite particles 
onto the Padova isochrones (\cite{Bertelli08};\cite{Bertelli09}).

\section{Results}
\label{res}

For this pilot study, we situate the observer at an analogous `solar 
neighbourhood', much as we did in \cite{Gibson13}.  From that vantage 
point, we `view' the simulated galaxy, first isolating composite stellar 
particles within $d$=10~kpc of the observer and which avoid the 
galactic plane (i.e., enforcing a galactic latitude restriction of 
$|b|$$>$20$^{~\circ}$ from the observer's position).  Such galactic plane 
restrictions mimic the typical constraint imposed upon ground-based, 
optically-selected, all-sky surveys, such as RAVE \cite{Steinmetz06}.

\begin{figure}[ht] 
\centering
\begin{tabular}{cc}
\hspace*{3.5cm}\includegraphics[width=80mm]{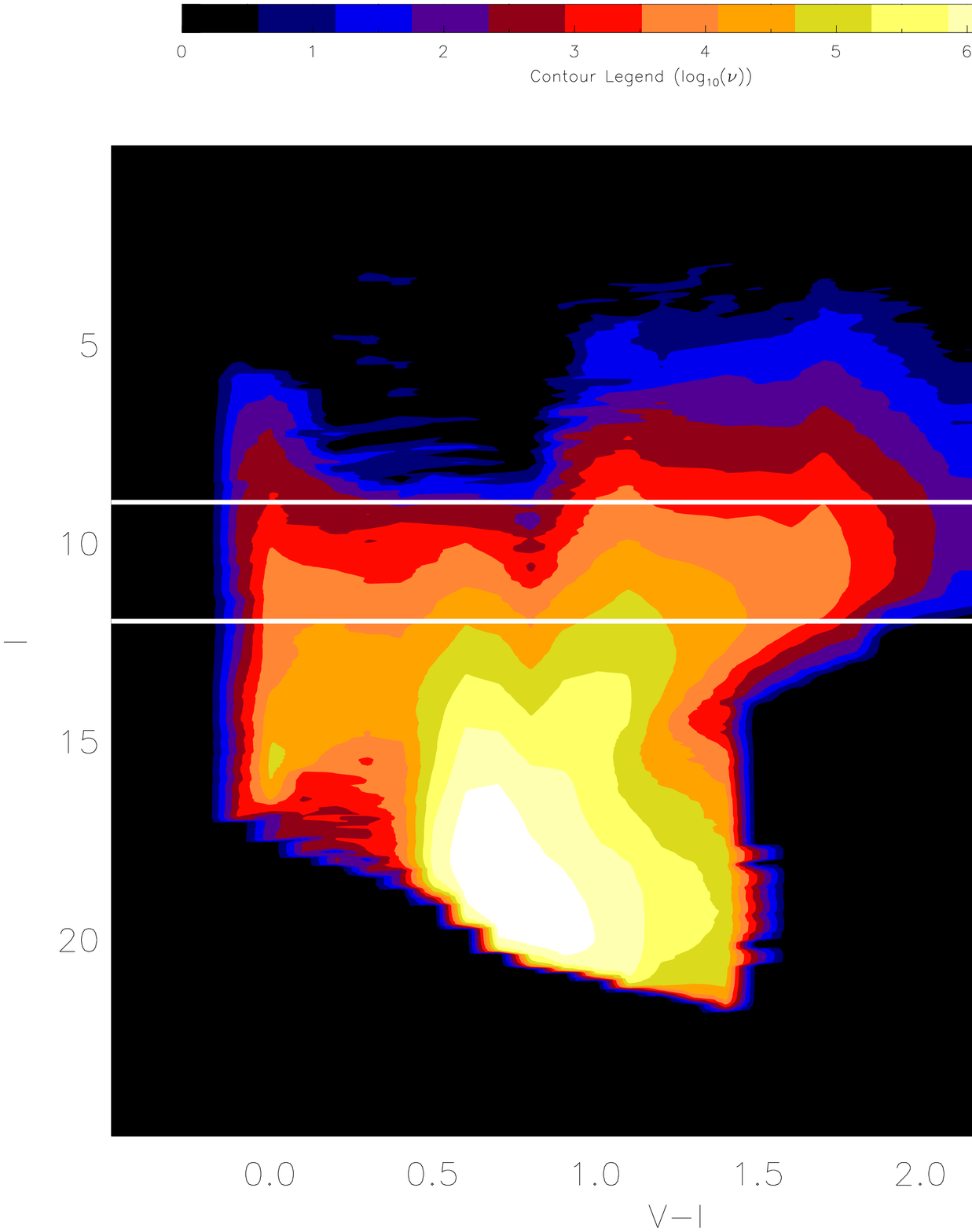}\\
\hspace*{-3.5cm}\includegraphics[width=80mm]{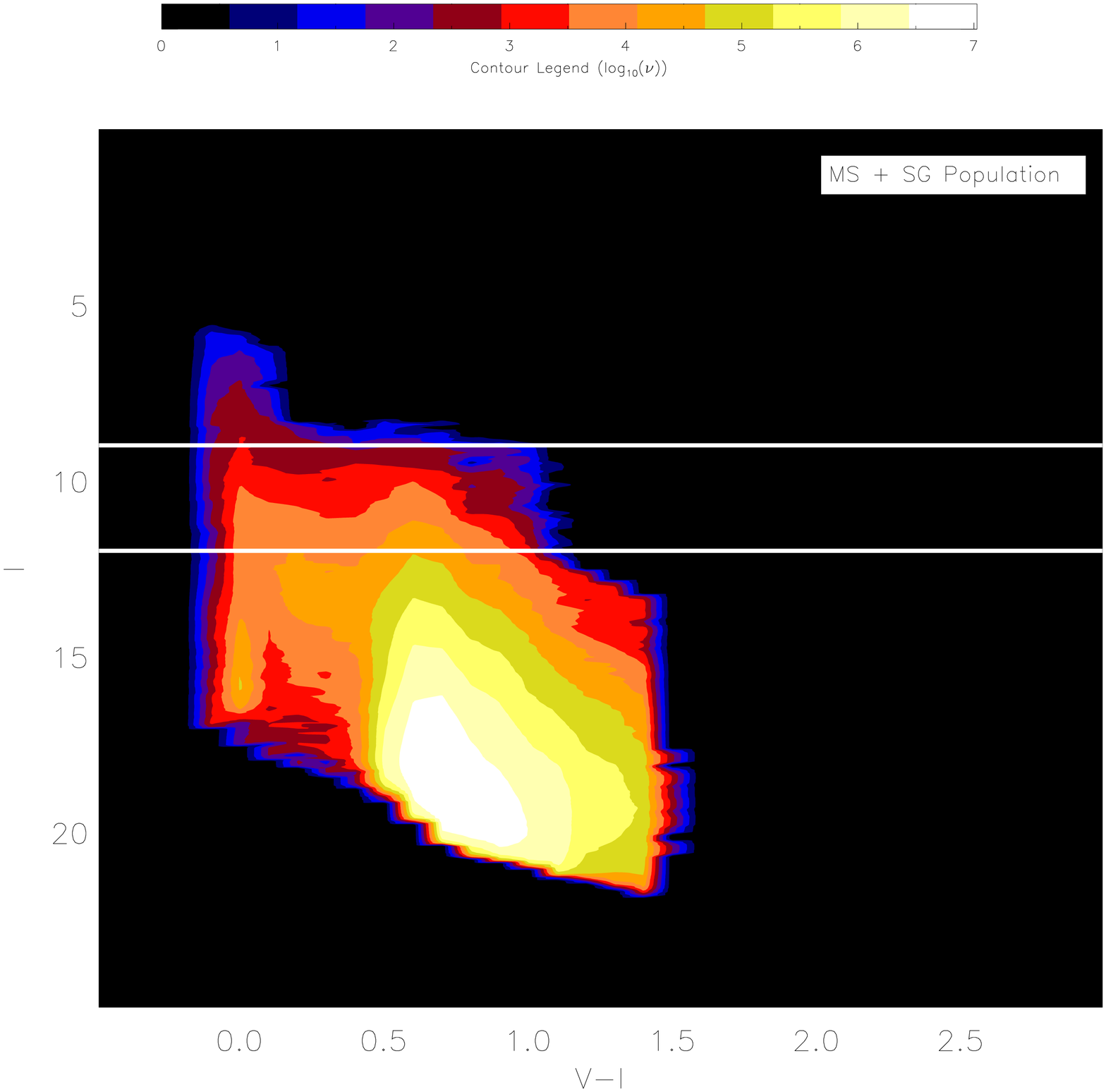}&
\hspace{-4.5cm}\includegraphics[width=80mm]
{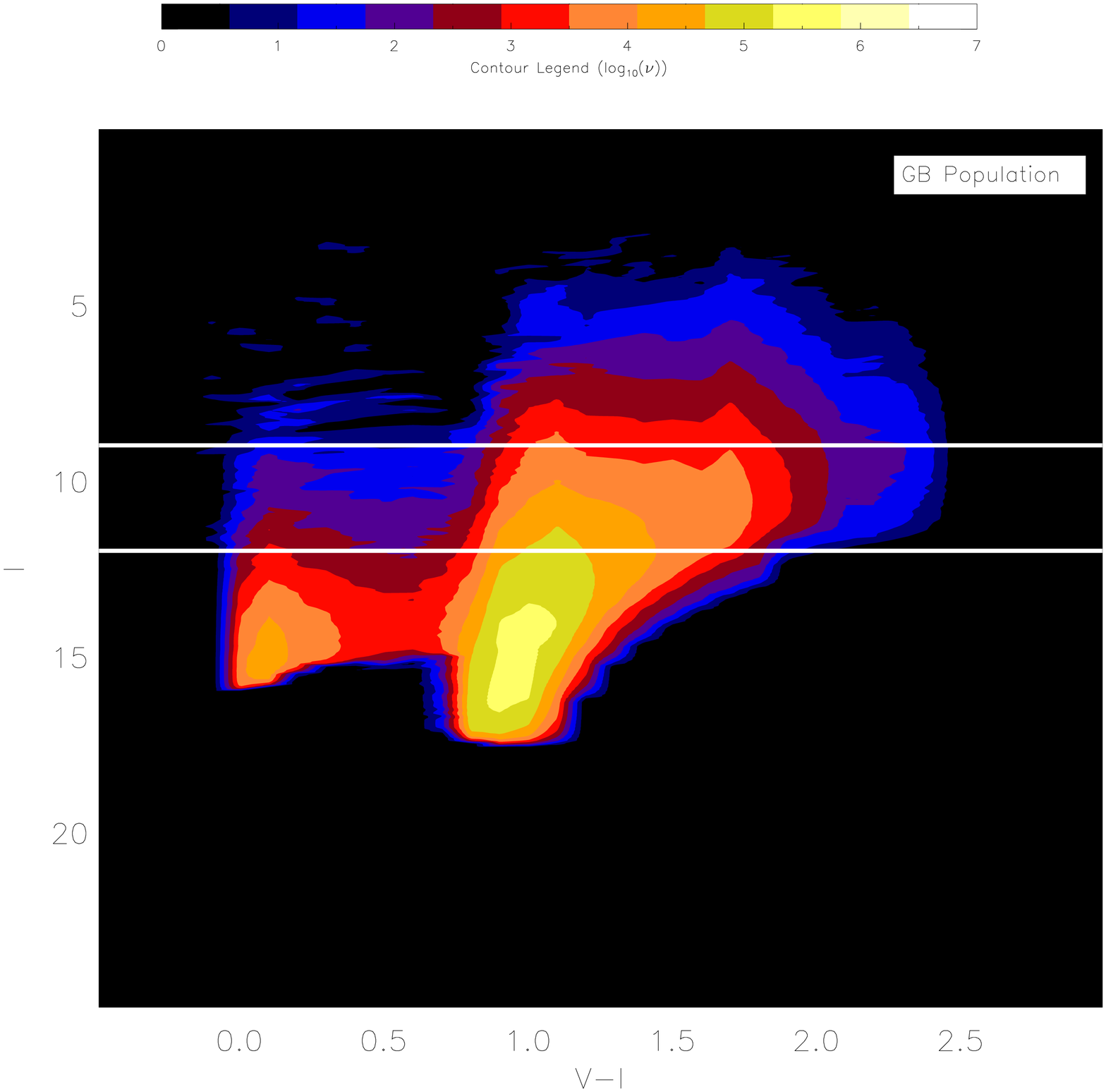}
\end{tabular}
\caption{Upper: CMD for the 
synthetic stars passing the spatial cut, 
colour-coded by density. The two 
lines represent our I-band selection cut.
Bottom left: As for the upper, but 
including gravity cut for main sequence + sub-giants.
Bottom right: As for the upper, but 
including gravity cut for giant branch stars.}
\label{cmd}
\end{figure}

For each composite stellar particle passing the spatial cut, we employ 
\textsc{SynCMD} to distribute the underlying $\sim$10$^{~5}$ individual 
`synthetic' stars, in the appropriate numbers (weighted by the initial 
mass function), into each `bin' of the associated colour-magnitude 
diagram (CMD). After taking into account the distance modulus of each 
such star, the relevant colour-apparent magnitude diagram, from the 
observer's vantage point, results.  The upper panel of Fig~\ref{cmd} 
shows the resulting (I,V$-$I) CMD for the aforementioned stars passing 
the initial spatial cut; colour-coding here is by density of stars 
populating each CMD bin, as noted in the legend.

Having transformed our composite stellar particles into individual 
synthetic stars, each with colours, luminosities, and gravities drawn 
from the Padova isochrones, we are now in a position to apply 
observationally-motivated selection criteria to the simulation. For this 
pilot study, we employ criteria not dissimilar to that of RAVE 
\cite{Steinmetz06}.  We will show the results of three experiments: (i) 
including all stars with an I-band magnitude in the range 9$<$$I$$<$12, 
regardless of gravity, which is reflected in the white horizontal lines 
in the upper panel of Fig~\ref{cmd}; (ii) including both the 
aforementioned I-band selection, and a surface gravity cut aimed at 
isolating main sequence and sub-giant branch (MS+SG) stars ($\log 
g$$>$3.5); and (iii) including both the aforementioned I-band selection, 
and a surface gravity cut aimed at isolating giant branch (GB) stars 
($\log g$$<$3.5).  The bottom left and right panels of Fig~\ref{cmd} 
correspond to the colour-apparent magnitude diagrams associated with 
(ii) and (iii), respectively.

To assess the impact of our observationally-motivated
selection criteria in a quantitative sense, for this pilot study
we examine the resulting metallicity distribution function (MDF) 
for each of the aforementioned \textsc{SynCMD} (i.e., `synthetic') 
experiments (referred to as (i), (ii), and (iii) in the previous 
paragraph), and contrast those with the MDF inferred from using
only the `composite' stellar particles.

\begin{figure}[ht]
\begin{center}
\hspace{0.25cm}
\psfig{figure=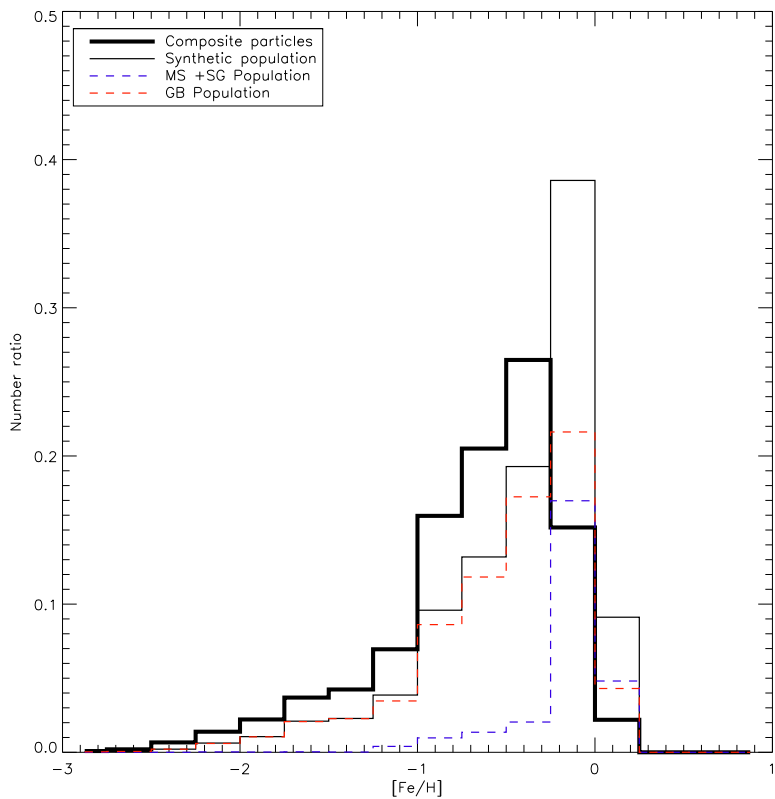,width=11.cm}
\caption{Comparative metallicity distribution function of the composite 
particles from our simulated galaxy (thick black line) and results from 
the decomposition into individual stars made with \textsc{SynCMD} (thin 
black line). We have further decomposed this synthetic population into 
main sequence + sub-giants (MS+SG, blue dashed line) and giant branch 
stars (GB, red dashed line), according to the $\log g$ restrictions 
described in \S~3.}
\label{mdf}
\end{center}
\end{figure}

The resulting four MDFs are shown in Fig~\ref{mdf}: the thick solid line 
corresponds to the sub-set of `composite' stellar particles passing the 
initial spatial cut, while the thin solid line corresponds to the case 
in which also an I-band cut has been applied.  One can 
immediately appreciate that the distributions, while similar globally, 
differ in detail; formally, the MDF inferred from the composite 
simulation particles is significantly less skewed, with significantly 
lower kurtosis, than the counterpart derived employing a RAVE-like 
I-band selection cut (skewness=$-$1.2 vs $-$1.5; kurtosis=1.4 vs 2.5). 
The differences become even more obvious when further constraining the 
observational cuts beyond just apparent magnitude, to isolate main 
sequence plus sub-giant branch (MS+SG) stars (blue-dashed histogram in 
Fig~\ref{mdf}) or giant branch (GB) stars (red-dashed histogram in 
Fig~\ref{mdf}). In those cases, the skewness ranges from $-$2.2 (MS+SG) 
to $-$1.3 (GB), while the kurtosis ranges from 7.1 (MS+SG) to 1.7 (GB).  
Differences in the mean and variance also range from 0.1 to 0.2~dex, 
across the four experiments.

In and of themselves, these quoted moments of the MDFs might not be 
enlightening, but a comparison with the corresponding moments from the 
broad suite of simulations described by \cite{Pilkington12} proves 
interesting.  In \cite{Pilkington12}, five variants of the same 
simulated disk galaxy were presented, employing different initial mass 
functions, different treatments of radiation pressure from massive 
stars, different treatments of metal diffusion, and different treatments 
of the thermal cooling timescale.  These five simulations were identical 
in every other consideration; as Table~2 in \cite{Pilkington12} shows, the 
resulting ranges in `solar neighbourhood' MDF skewness and kurtosis vary 
from $-$0.9 to $-$1.8 (skewness), and 0.9 to 3.8 (kurtosis).
\it In other words, the impact of simply how one observes a simulation 
(i.e., observationally-motivated, as we have done here, or simple 
spatial cuts, as has been done previously by essentially the entire 
field) can be as quantitatively important as any of the sub-grid physics 
treatments within the simulations themselves!  \rm

In some sense, the results shown here should not be shocking, and are 
already appreciated by the observational community.  For example, for 
the bright apparent magnitude cut employed here, the main sequence and 
sub-giant branch stars all lie within 1$-$2~kpc of the simulation 
observer, while the intrinsically brighter giants probe much greater 
distances.  Without an \it a posteriori \rm matching volume-limited 
selection criterion for the two samples (MS+SG vs GB), the different 
effective distances of the samples mean that different metallicities 
and ages are encountered due to the interplay between the 
radial+vertical metallicity gradients and the spatially varying 
age-metallicity relations.

\section{Summary}

We have transformed a cosmological hydrodynamical simulation of an
$\sim$L$\ast$ Milky Way-like galaxy from the simulator's space to the
observer's plane, making use of the synthetic colour-magnitude diagram
tool.  Having replaced $\sim$10$^{~6}$ `composite' stellar particles with
$\sim$10$^{~11}$ `synthetic' stars, we have employed various spatial,
apparent magnitude, and surface gravity cuts to the individual 
synthetic stars, and viewed the simulation much as an observer would 
do in nature.  This pilot study was not meant to be exhaustive, but
instead, to illustrate that under very common conditions, selecting 
stars based upon actual observable characteristics, rather than just 
using massive composite simulated stellar particles, can lead to 
inferred metallicity distributions which differ by as much as any of the
various sub-grid physics treatments employed by simulators.  While 
an enormous literature and community has evolved which is predicated
upon understanding the subtleties of these sub-grid physics
implementations, we have shown that it is equally as important to
simply view the simulation `correctly' as it is to get the 
sub-grid physics `right'.  In the next phase of our work, we 
will also include the effect of foreground reddening, as well
as explore the impact of magnitude-limited vs volume-limited
sample selection upon not only the MDF, but also various 
chemistry-chemistry projections, the age-metallicity relation, 
and metallicity gradient determinations.

\end{document}